\documentclass[CEJP,PDF]{cej} 
\usepackage{layout}
\usepackage{amsmath}
\usepackage{textcomp}
\usepackage{hyperref}

\title{New results from fluctuation analysis in NA49 at the CERN SPS}

\articletype{Research Article}
\author{Maja Ma\'{c}kowiak-Paw{\l}owska\inst{1}\inst{2}\email{majam@if.pw.edu.pl} for the NA49 Collaboration,}
\institute{
     \inst{1} Goethe University, Frankfurt,\\
     Max-von-Laue-Str. 1, 60438 Frankfurt am Main, Germany
     \inst{2} Warsaw University of Technology,\\
     ul. Koszykowa 75, 00-662 Warszawa, Poland
          }

\abstract{The exploration of the phase diagram of strongly interacting matter, particularly the study
of the phase transition from hadronic to partonic matter
and the search for a hypothetical critical endpoint of the first order transition line, is one of the most challenging
tasks in present heavy ion physics.\newline
\indent In this talk new results on chemical (particle ratio), transverse momentum, multiplicity and azimuthal angle fluctuations will be presented.
We also discuss their connection to the onset of deconfinement and to the critical endpoint.}



\begin{document}

\maketitle


\vspace{-1cm}
\section{Introduction}

The NA49 experiment studies an important region of the phase diagram of strongly interacting matter. First, the Statistical Model of the Early Stage (SMES) of nucleus-nucleus collisions \cite{marek} predicted the energy threshold for deconfinement at low SPS energies. Several structures in the excitation functions were expected within the SMES: a kink in the pion yield per participant nucleon (change of slope due to increased entropy as a consequence of the activation of partonic degrees of freedom), a sharp peak (horn) in the strangeness to entropy ratio, and a step in the inverse slope parameter of transverse mass spectra (constant temperature and pressure in a mixed phase). Such signatures were observed in central Pb+Pb collisions by the NA49 experiment around $\sqrt{s_{NN}}=7.6$~GeV~\cite{OD}. Fluctuation analysis may provide additional evidence of the onset of deconfinement. \newline
\indent Second, lattice QCD calculations suggest a critical point of strongly interacting matter which may be observable in the SPS energy range~\cite{fodor}. Fluctuations and correlations are basic tools to study this phenomenon. We expect enlarged fluctuations close to the critical point. In nucleus-nucleus collisions a maximum of fluctuations is expected when freeze-out happens near the CP.

\section{NA49 results on fluctuations}
\subsection{Particle ratio fluctuations}
NA49 uses the $\sigma_{dyn}$ quantity to measure particle ratio fluctuations. From event-by-event particle ratio distributions ($A/B$, e.g. $K/\pi$) for data and mixed events the quantity $\sigma$ is defined as:
\begin{equation}
\sigma=\frac{\sqrt{Var(A/B)}}{<A/B>}\cdot100\%~,
\end{equation}
where $<A/B>$ and $Var(A/B)$ are mean and variance of the distribution, respectively.

\begin{figure}
\includegraphics[width=0.3\textwidth]{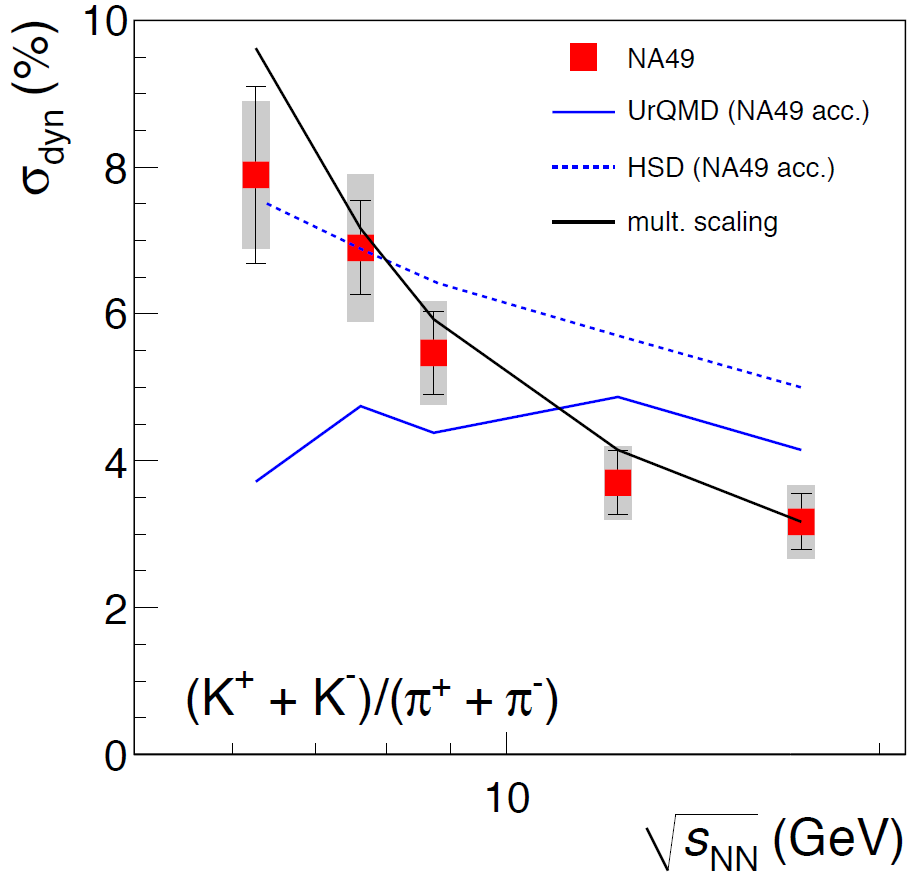}
\includegraphics[width=0.3\textwidth]{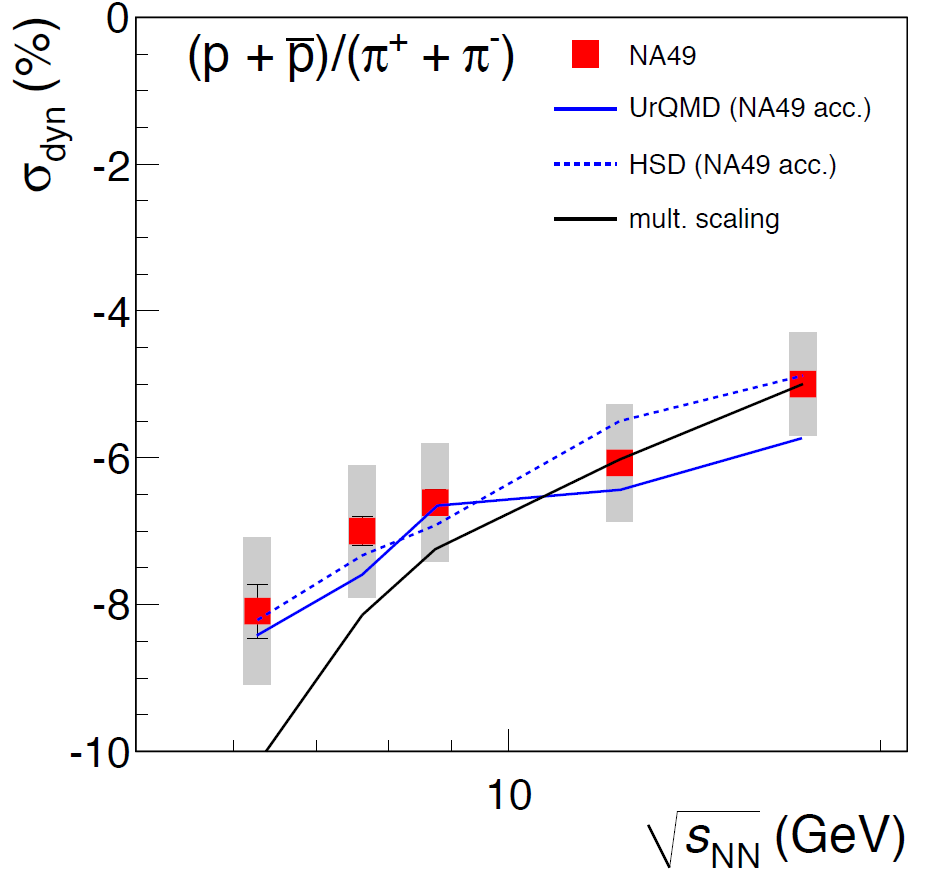}
\includegraphics[width=0.3\textwidth]{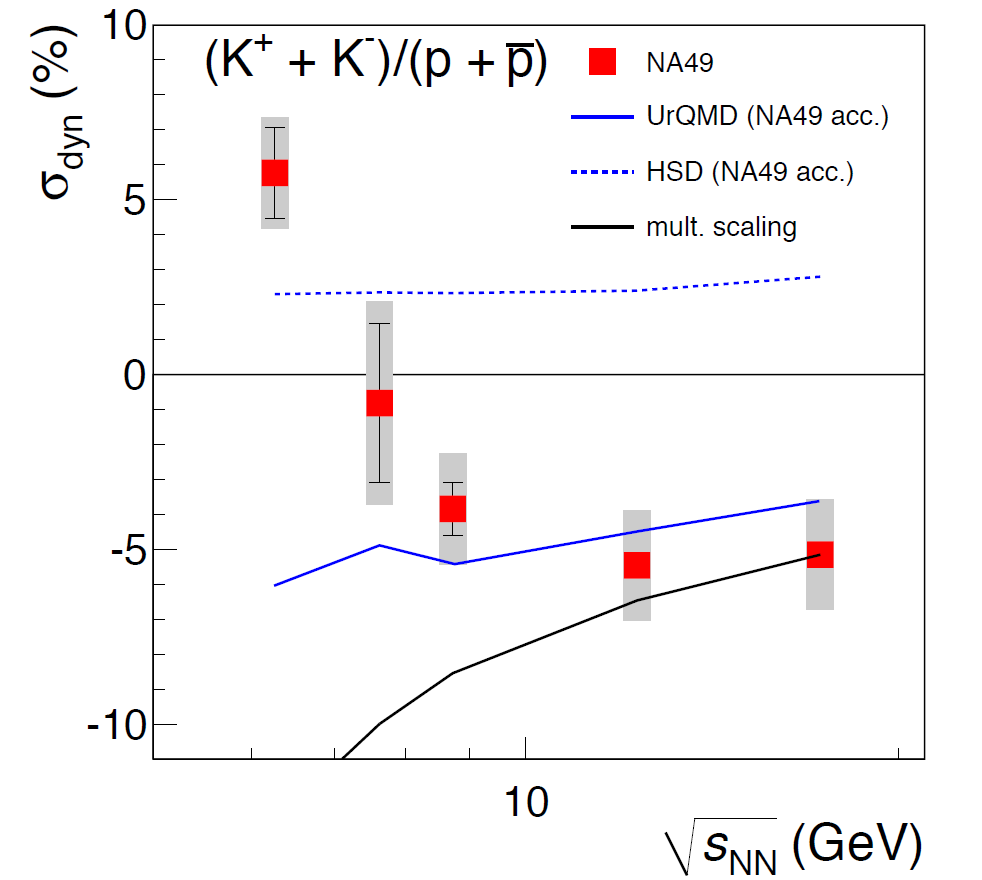}
\vspace{-0.6cm}
\caption{Energy dependence of the $K/\pi$, $p/\pi$ and $K/p$ ratio fluctuations in central Pb+Pb collisions measured by NA49~\cite{tim}. Model predictions are indicated by lines.}
\label{sigma1}
\end{figure}

The fluctuation measure $\sigma_{dyn}$ is then calculated as:
\begin{equation}
\sigma_{dyn}=sign(\sigma_{data}^{2}-\sigma_{mixed}^{2})\sqrt{|\sigma_{dyn}^{2}-\sigma_{mixed}^{2}|}~,
\end{equation}
where subscripts "data" and "mixed" refer to $\sigma$ calculated for the data and mixed event distributions, respectively.
A detailed description of this method can be found in Ref. \cite{tim}.\newline
The energy dependence of event-by-event fluctuations of particle ratios in the $3.5\%$ most central Pb+Pb collisions is presented in Fig. \ref{sigma1}.
For $K/\pi$ $\sigma_{dyn}$ is positive and shows a rise towards low SPS energies which is not reproduced by the UrQMD model. The HSD model catches the trend but over-predicts measurements at high SPS energies. Data points are reproduced by the "Poisson" multiplicity scaling proposed in Ref. \cite{volker}. For $p/\pi$ $\sigma_{dyn}$ is negative and shows a decrease towards low SPS energies which is reproduced by the hadronic models and the multiplicity scaling. The trend is understood in terms of correlations due to nucleon resonance decays.\newline
In case of $K/p$, $\sigma_{dyn}$ changes sign. This behavior cannot be reproduced neither by the hadronic models nor by the multiplicity scaling. The centrality dependence of $\sigma_{dyn}$ for Pb+Pb interactions at $\sqrt{s_{NN}}=17.3$ GeV (see figures in conference slides) agrees with the UrQMD predictions and the multiplicity scaling for $K/\pi$ and $p/\pi$. For $K/p$ the UrQMD and the multiplicity scaling slightly overpredict $\sigma_{dyn}$ values. For details see Ref. \cite{tim}.\newline
Recently the STAR Collaboration has presented results on the particle ratio fluctuations using the $\nu_{dyn}$ measure \cite{terry}. A comparison between NA49 and STAR results is done using an approximate relation between $\sigma_{dyn}$ and $\nu_{dyn}$: $\nu_{dyn}\approx sign(\sigma_{dyn})\sigma_{dyn}^{2}$. For $K/\pi$ and $K/p$ STAR results do not show an increase towards low SPS energies. Analysis procedures were checked by both collaborations and no problems have been found so far. The difference may be due to different acceptances and centrality selection procedures. Both experiments correct their results for incomplete particle identification only approximately. In order to eliminate this source of systematic bias, the NA49 experiment will use a new analysis method, the so-called identity method \cite{id}. 

\subsection{Multiplicity and average transverse momentum fluctuations}
\begin{figure}
\includegraphics[width=0.3\textwidth]{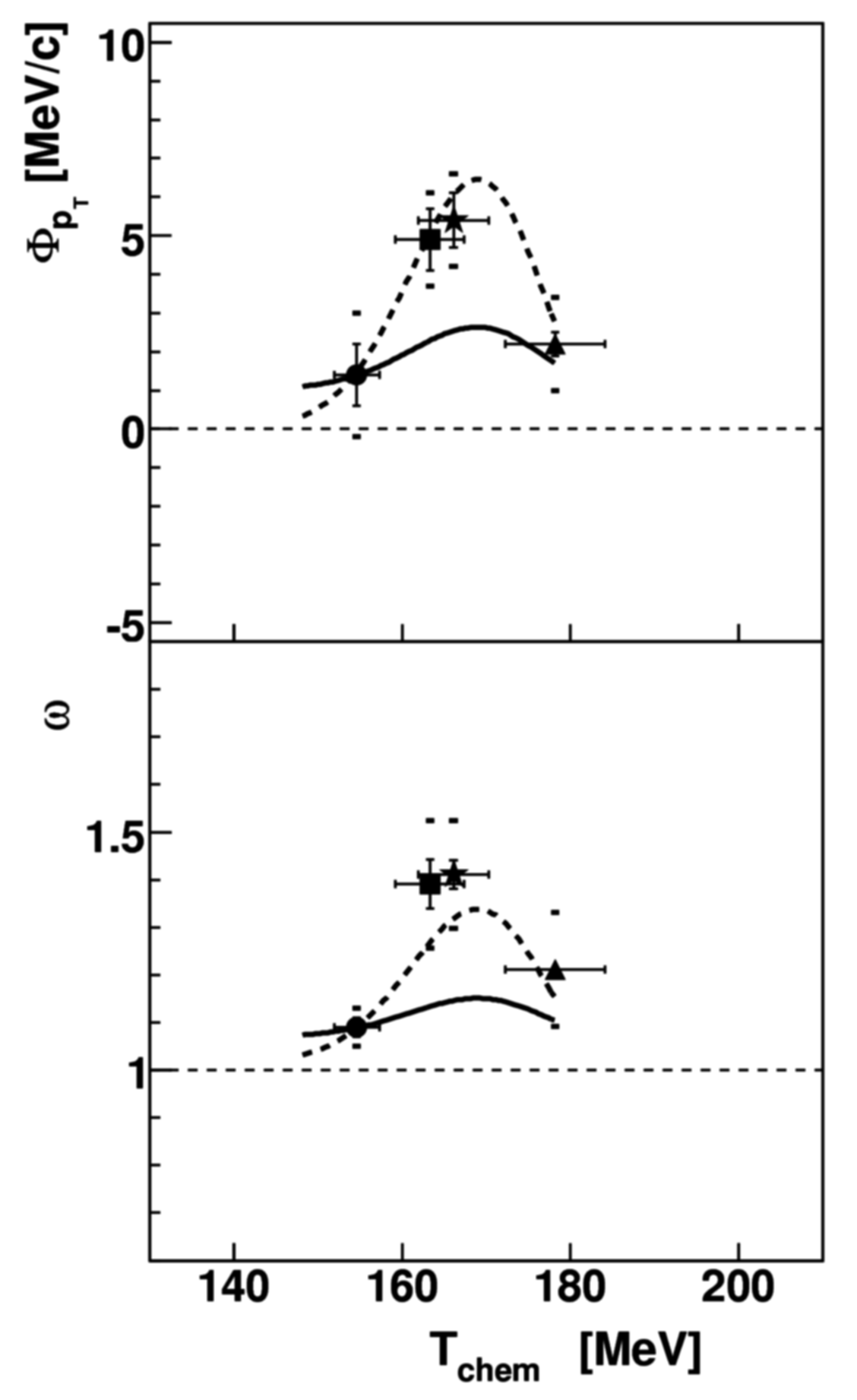}
\includegraphics[width=0.3\textwidth]{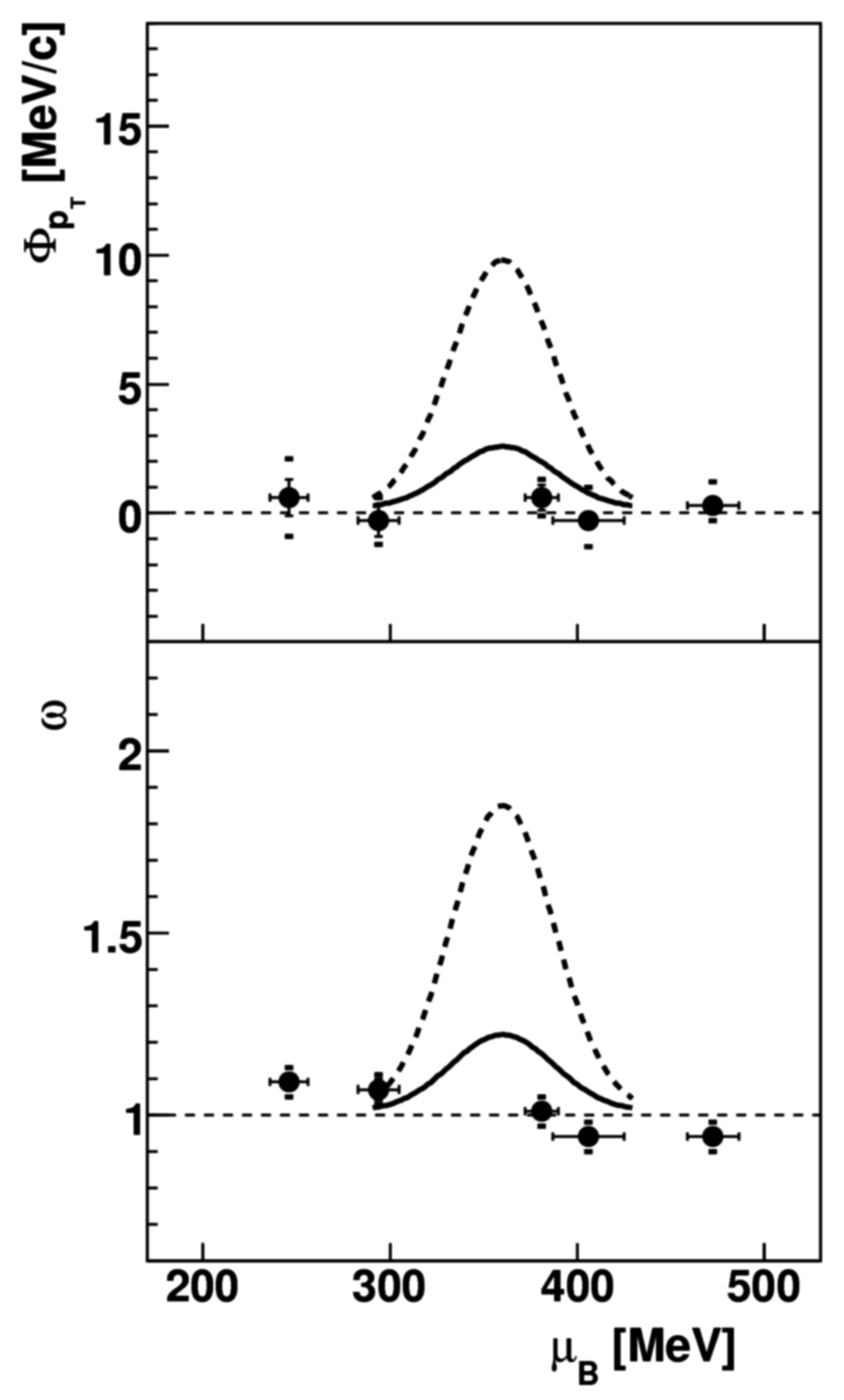}
\vspace{-0.6cm}
\caption{System size ($T_{chem}$) dependence at $\sqrt{s_{NN}}=17.3$ GeV and energy ($\mu_{B}$) dependence of $\Phi_{p_{T}}$ and $\omega$ \cite{kasiaSQM}.  Lines correspond to the critical point predictions (see the text and \cite{kasiaSQM} for details).}
\label{fipt}
\end{figure}

Enhanced fluctuations of multiplicity and mean transverse momentum were suggested as an important signature of the critical point \cite{stephanov}. In order to quantify multiplicity fluctuations NA49 uses the scaled variance of the multiplicity distribution ($\omega=\frac{<N^{2}>-<N>^{2}}{<N>}$)\cite{na49omega,na49omega2}. This is an intensive measure, which for a Poisson distribution is equal to 1. Results on $\omega$ are presented for the $1\%$ of most central collisions.
Transverse momentum fluctuations were measured by the NA49 experiment using the $\Phi_{p_{T}}$ quantity \cite{na49fipt,na49fipt2}.\newline
Dependence on system size at $\sqrt{s_{NN}}=17.3$~GeV and energy for central Pb+Pb collisions is shown in Fig.~\ref{fipt}. For each reaction the chemical freeze-out temperature ($T_{chem}$) and baryonic chemical potential ($\mu_{B}$) were taken from the fits of Ref.~\cite{Becattini:2005xt}.
Fig.~\ref{fipt} presents the system size and energy dependences versus $T_{chem}$ and $\mu_{B}$, respectively. The lines correspond to predictions for the critical point for $\Phi_{p_{T}}$ and $\omega$, for details see Ref. \cite{kasiaSQM}. The data show weak, if any, energy ($\mu_{B}$) dependence. For system size ($T_{chem}$) dependence at the top SPS energy NA49 finds a maximum of $\Phi_{p_{T}}$ and $\omega$ for C+C and Si+Si collisions. It agrees with predictions for the critical point located at $\mu_{B}=250$ MeV, $T_{chem}=178$~MeV.  Moreover, the signal for negatively charged particles is two times weaker than for all charged particles as expected for fluctuations due to the critical point. For more details see Ref.~\cite{kasiaSQM}.  

\subsection{Azimuthal angle fluctuations}
\begin{figure}
\includegraphics[width=0.4\textwidth]{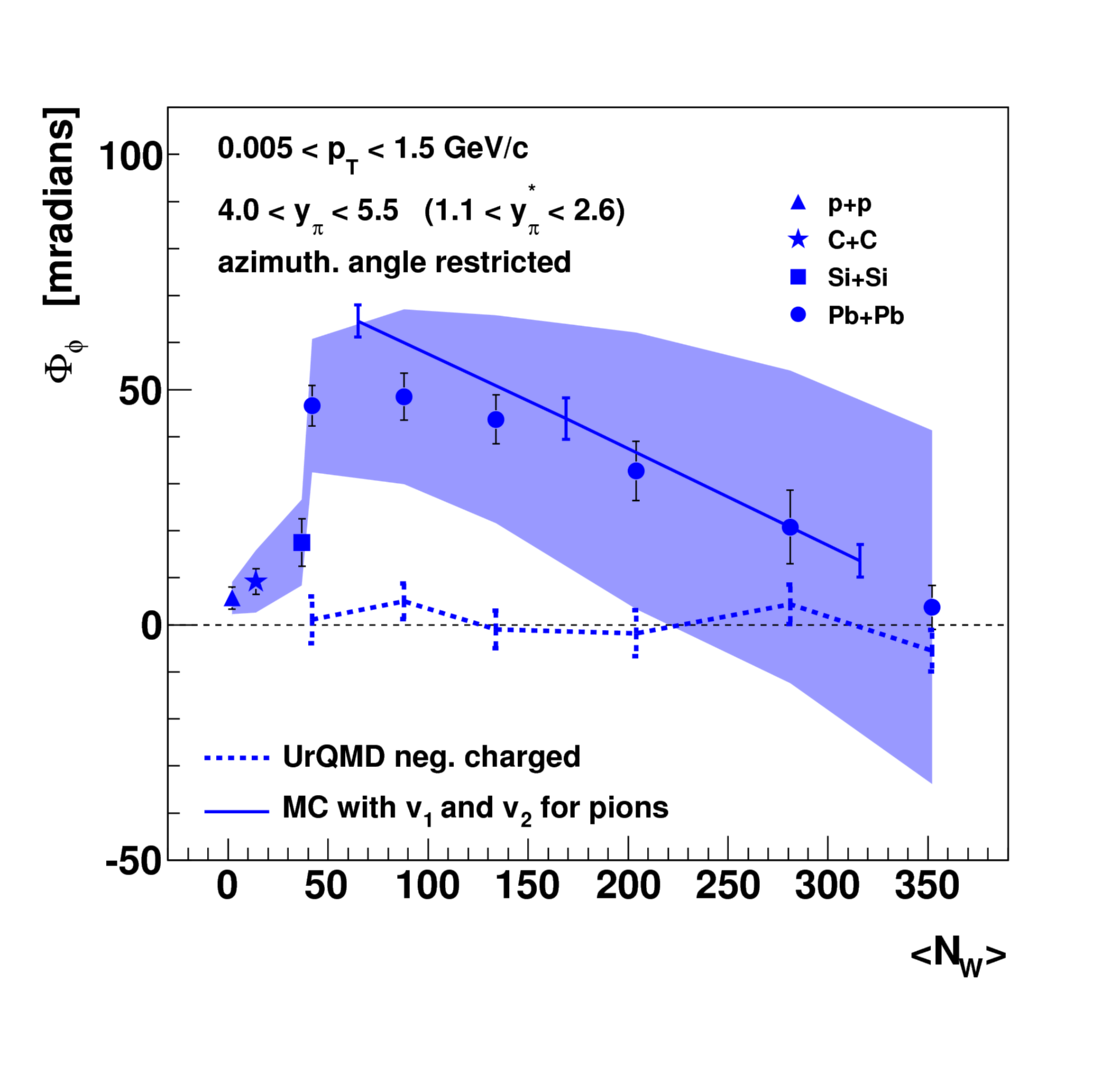}
\includegraphics[width=0.4\textwidth]{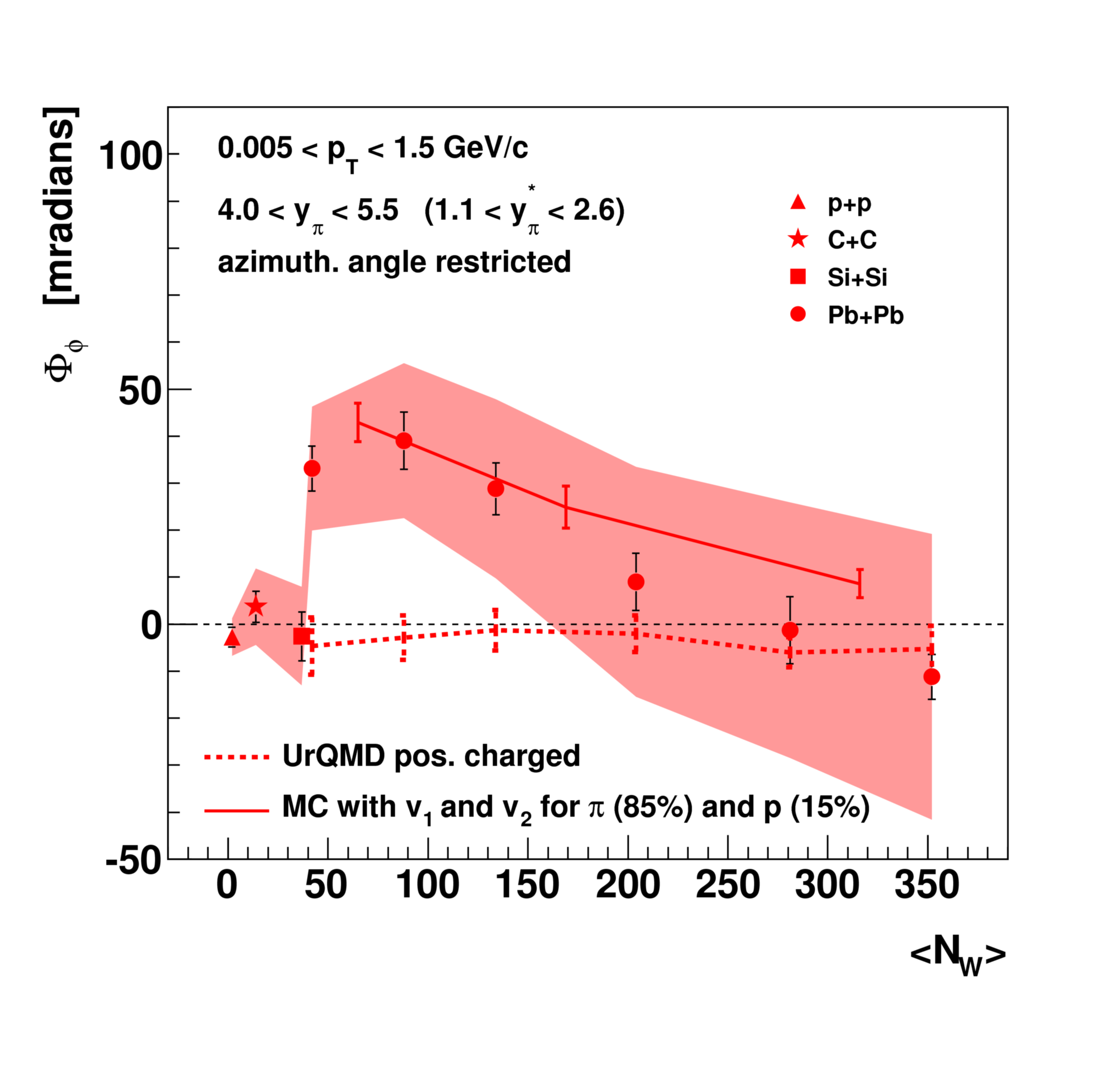}
\vspace{-0.6cm}
\caption{System size dependence of azimuthal fluctuations in Pb+Pb collisions at $\sqrt{s_{NN}}=17.3$~GeV. NA49 data and predictions are evaluated in the same kinematical acceptance defined in Ref. \cite{na49fipt}) for Pb+Pb collisions at $\sqrt{s_{NN}}=17.3$~GeV. Results are preliminary.}
\label{fifi}
\end{figure}

Azimuthal angle fluctuations are believed to be sensitive to plasma instabilities~\cite{plasma} and flow fluctuations~\cite{flow,flow2}. Besides those effects, there are several other background effects, which may influence azimuthal angle fluctuations, among them are resonance decays, momentum conservation, flow, (di-)jets and quantum statistics. The NA49 experiment uses the $\Phi_{\phi}$ measure to study those fluctuations. \newline
For central Pb+Pb interactions there is no energy ($\mu_{B}$) dependence for positively and negatively charged particles. $\Phi_{\phi}$ is positive for negatively charged particles and its value is not reproduced by the UrQMD model. For positively charged particles $\Phi_{\phi}$ is consistent with zero and it agrees with the model predictions \cite{fifi2}. System size ($T_{chem}$) dependence of $\Phi_{\phi}$ is presented in Fig. \ref{fifi} for negatively (left) and positively (right) charged particles. $\Phi_{\phi}$ is positive, reaches maximum for peripheral Pb+Pb interactions and decreases for more central ones. This behavior is not reproduced by the UrQMD model but the magnitude of $\Phi_{\phi}$ is reproduced by a simple Monte Carlo model with directed ($v_{1}$) and elliptic ($v_{2}$) flows included~\cite{fifi1}.

\section{Summary}
Energy and system size dependence of $K/\pi$ and $p/\pi$ fluctuations can be described in the simple multiplicity scaling model. The $K/p$ fluctuations for central Pb+Pb collisions change sign at the low SPS energies and show qualitative deviations from this scaling as well as from the predictions of the HSD and UrQMD models. STAR results for $K/\pi$ and $K/p$ fluctuations in central Au+Au interactions indicate a different energy dependence than the one measured by NA49. Both collaborations continue to work on the clarification of the observed differences. \newline
Fluctuations of multiplicity and transverse momentum are maximal for Si+Si at the top SPS energy~($\sqrt{s_{NN}}=17.3$~GeV). This, together with a maximal intermittent behavior of di-pions and protons \cite{Anticic:2009pe,fotis}, might be connected with the critical point. The results strongly motivate future experiments in the CERN SPS energy range.  \newline
Azimuthal angle fluctuations show no energy dependence. $\Phi_{\phi}$ is maximal for peripheral Pb+Pb collisions at the top SPS energy and decreases for more central collisions.  This behavior can be understood as the effect of directed and elliptic flow.

\vspace{0.5cm}
\small{\textbf{Acknowledgments}: This work was partially supported by Polish Ministry of Science and Higher Education under grant N N202 204638 and the German Research Foundation (grant GA 1480/2.1).}

\end{document}